\documentclass[twocolumn,showpacs,preprintnumbers,superscriptaddress,amsmath,amssymb,floatfix,prl]{revtex4}

\usepackage{graphicx}
\usepackage{dcolumn}
\usepackage{bm}
\usepackage{soul}

\usepackage[utf8]{inputenc}
\usepackage{xcolor}

\newcommand{\vkO}{\overrightarrow{k}_0}
\newcommand{\vkp}{\overrightarrow{k}_{\textrm{p}}}
\newcommand{\vkSRS}{\overrightarrow{k}_{\textrm{SRS}}}
\newcommand{\vkReSRS}{\overrightarrow{k}_{\textrm{Re-SRS}}}
\newcommand{\vkpRe}{\overrightarrow{k}_{\textrm{p}}'}

\newcommand{\wSRS}{\omega_{\textrm{SRS}}}
\newcommand{\wReSRS}{\omega_{\textrm{Re-SRS}}}

\begin{document}

\title{Experimental evidence of stimulated Raman re-scattering in laser-plasma interaction}

\author{J.-R. Marquès}
\email[]{jean-raphael.marques@polytechnique.fr}
\affiliation{LULI, CNRS, CEA, Sorbonne Université, École Polytechnique, Institut Polytechnique de Paris, 91128 Palaiseau, France}

\author{F. Pérez}
\affiliation{LULI, CNRS, CEA, Sorbonne Université, École Polytechnique, Institut Polytechnique de Paris, 91128 Palaiseau, France}

\author{P. Loiseau}
\affiliation{CEA, DAM, DIF, F-91297 Arpajon, France}
\affiliation{Université Paris-Saclay, CEA, LMCE, 91680 Bruyères-le-Châtel, France}

\author{L.~Lancia}
\affiliation{LULI, CNRS, CEA, Sorbonne Université, École Polytechnique, Institut Polytechnique de Paris, 91128 Palaiseau, France}

\author{C. Briand}
\affiliation{LESIA, Observatoire de Paris-PSL, CNRS, Sorbonne Université, Université de Paris, 92195 Meudon, France}

\author{S. Depierreux}
\affiliation{CEA, DAM, DIF, F-91297 Arpajon, France}

\author{M. Grech}
\affiliation{LULI, CNRS, CEA, Sorbonne Université, École Polytechnique, Institut Polytechnique de Paris, 91128 Palaiseau, France}

\author{C. Riconda}
\affiliation{Sorbonne Université, LULI, CNRS, CEA, École Polytechnique, Institut Polytechnique de Paris, 75255 Paris, France}

\date{\today}

\begin{abstract}
We present the first experimental evidence of stimulated Raman re-scattering of a laser in plasma: The scattered light produced by the Raman instability is intense enough to scatter again through the same instability. Although never observed, re-scattering processes have been studied theoretically and numerically for many years in the context of inertial confinement fusion (ICF), since the plasma waves they generate could bootstrap thermal electrons to high energies [Phys. Rev. Lett. \textbf{110}, 165001 (2013)], preheating the fuel and degrading ignition conditions. Our experimental results are obtained with a spatially smoothed laser beam consisting of many speckles, with an average intensity around $10^{14}$ W/cm$^2$ and close to $10^{15}$ W/cm$^2$ in the speckles, such as those usually used in direct-drive ICF. Kinetic and hydrodynamic simulations show good agreement with the observations.
\end{abstract}

\maketitle

Parametric laser-plasma instabilities (LPIs)\cite{Kruer,Michel}, such as stimulated Raman scattering (SRS) or stimulated Brillouin scattering (SBS), are of fundamental interest in many fields involving the propagation of an intense laser pulse in an underdense plasma. These processes have long been studied in inertial confinement fusion for several reasons, both for laser indirect drive and direct drive: The loss due to backscattering, which reduces the amount of laser energy coupled to the target; the energy exchange (via SBS) between laser beams\cite{Kirkwood1996,Strozzi2017}, which can impact implosion by introducing drive asymmetries; or the excitation (via SRS) of electron plasma waves (EPW) which, through the associated electric fields, can trap and accelerate thermal electrons to higher energies~\cite{Dewald2010,Christopherson,Winjum2013}. These energetic (hot) electrons can preheat the fuel~\cite{Dewald2016} and prevent compression of the capsule to ignition conditions.

On the indirect drive side, the first National Ignition Facility (NIF) experiments showed electron heating to energies above 100 keV~\cite{Dewald2010}. While the low temperature part ($T_e\sim$ 10-20 keV) of the hot electron distribution was attributed to backward SRS, the phase velocity of the associated EPW was too low to produce the high temperature part. Using particle-in-cell (PIC) simulations, Winjum et al.~\cite{Winjum2013} showed that electrons can be progressively heated as they move between waves of increasing phase velocity produced by SRS rescattering (Re-SRS). In SBS and SRS, the incident laser excites and scatters low-frequency ion acoustic waves and high-frequency EPWs, respectively. If the incident laser is intense enough, the daughter (scattered) light wave may be intense enough to further excite a secondary SBS~\cite{Speziale1980,Montes1985,Langdon2002} or SRS~\cite{Langdon2002}. Design studies for NIF~\cite{Hinkel2004, Hinkel2005} showed from PIC simulations the possible occurrence of Brillouin backscattering of Raman forward scattering \cite{Hinkel2004}. Simulations by Langdon and Hinkel~\cite{Langdon2002} also showed that Raman scattering can be saturated by subsequent Brillouin and Raman rescattering. They noted that rescattering could alter the interpretation of experimental diagnostics as well as the distribution of laser energy between transmission into the target, scattering losses and generation of energetic electrons.
More recently, considering that the wavelength of scattered light of Stimulated Raman Backcattering (SRBS) in the gas-filled hohlraums is generally shorter than the simulated results based on the ray-tracing model~\cite{Hall2017,Strozzi2017}, Hao \textit{et al.}~\cite{Hao2021} argued that this could also be the result of Re-SRS. They showed that Re-SRS can act as a frequency filter of backscattered light of SRBS. Their investigation also suggests that Re-SRS could be a potential reason for the ‘energy deficit’ appearing in the gas-filled hohlraum experiments at NIF~\cite{MacLaren2014}.

With respect to direct drive, recent theoretical and numerical studies focused specifically on Raman rescattering of backward SRS (Re-SRBS) in large-scale inhomogeneous plasmas. Zhao \textit{et al.}~\cite{Zhao2019} showed that while the first-order convective SRS occurs in plasmas of electron density in the range [0.11, 0.2]$n_c$, its backscattered light can induce absolute instabilities by second-order rescattering of SRS within [0.0764, 0.11]$n_c$, with associated EPWs that can heat abundant electrons up to a few hundred keV. Here $n_c$(cm$^{-3}$) $\sim 1.1 \times 10^{21}/ \lambda_0^2  (\mu$m) is the critical density of incident light, with $\lambda_0$ its vacuum wavelength.

The above works were done at laser intensities of $10^{14}-10^{15}$ W/cm$^2$. Other ignition schemes such as shock ignition~\cite{Betti2007} and hybrid drive~\cite{He2016} use higher intensities, $\sim 10^{16}$ W/cm$^2$. In these context Ji \textit{et al.}~\cite{Ji2021} studied Re-SRS at moderately high laser intensities using a picosecond laser. From their PIC simulations, they found that Re-SRS could be considered a concern for higher intensity ignition scale regimes: i) the convective amplification of Re-SRBS causes depletion of its pump, the primary SRBS light, and may lead to an underestimation of the SRBS level in experiments, ii) the backward hot electrons are strongly correlated with the Re-SRBS modes and represent a significant energy loss, but one that is likely to be missed by the conventional hard x-ray diagnostics typically used on the higher density side of the plasmas.

Raman rescattering is not limited to backward SRS. In recent years, Raman sidescattering (SRSS), especially when propagating nearly perpendicular to the density gradient, has been widely observed at fusion facilities\cite{Depierreux2016,Michel2019,Glize2023,Hironaka2023}. In some cases, SRSS even dominates the backward SRS. The rescattering of SRSS (Re-SRSS) has been studied with two-dimensional PIC simulations \cite{Tan2024}. It appears that the EPW excited by Re-SRSS could rapidly trap and accelerate electrons perpendicular to the density gradient, but in a relatively small amount. In contrast to the rescattering of backscattered light, Re-SRSS does not allow any cascade acceleration effect, and thus does not produce hot-electron bursts as expected from Re-SRBS.

To the best of our knowledge, despite all these previous studies, rescattering of SBS or SRS has only been identified in simulations, with a presumably indirect signature given by the generation of very hot electrons. In this Letter, we present the first direct experimental evidence for stimulated Raman rescattering. These results are compared with PIC and hydrodynamic simulations.

\begin{figure}[t]
	\begin{center}
	\includegraphics[width=\columnwidth]{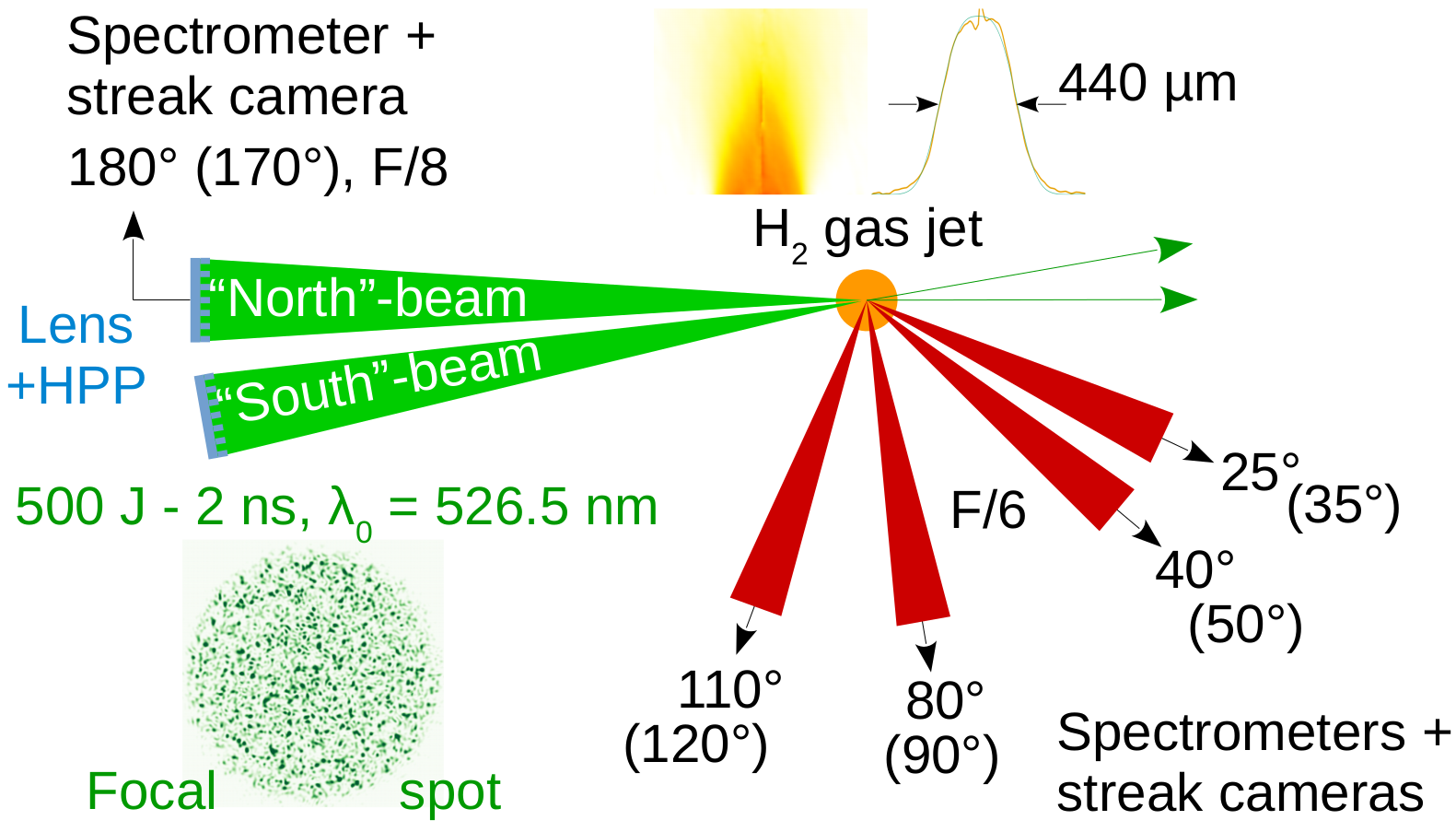}
	\caption{\small{\textit{Experimental setup. The "South"-beam propagates at 10$^\circ$ from the "North"-beam. The scattered light is collected at 5 angles. The angles in parenthesis are relative to the axis of the "South"-beam, the others to the axis of the "North"-beam.}}}
	\label{Setup}
	\end{center}
\end{figure}

The experiment was performed at the LULI2000 laser facility (École Polytechnique, France). The setup is shown in figure \ref{Setup}. We used either the "North"-beam or the "South" beam to irradiate a supersonic hydrogen gas jet of 0.5 mm diameter. The gas flows along the vertical axis, and the laser beams propagate in the horizontal plane, with the "South"-beam at 10$^\circ$ from the "North"-beam. Each 20 cm diameter beam delivers 500 J in a 2 ns-long laser pulse of $\lambda_0$ = 526 nm, and is focused by a lens of focal length 1.6 m coupled to a hybrid phase plate. The resulting focal spot (bottom-left in Fig. \ref{Setup}) has a speckle pattern forming a hyper-Gaussian spatial distribution with a full width at half maximum (FWHM) of $\sim$ 500 $\mu$m. The average intensity in the central region is $\sim 1.3\times 10^{14}$ W/cm$^2$, while the maximum speckle intensity is $\sim 6.5\times 10^{14}$ W/cm$^2$. The laser polarization is linear, with an angle of 45$^{\circ}$ with respect to the horizontal plane. The scattered light from the interaction region (gas jet center) is collected in this horizontal plane at 25$^\circ$, 40$^\circ$, 80$^\circ$ and 110$^\circ$ from the "North"-beam axis by silver-coated F/6 spherical mirrors, and at 180$^\circ$ with an aperture of F/8. For the shots using the "South"-beam, these angles correspond to scattering at 35$^{\circ}$, 50$^{\circ}$, 90$^{\circ}$, 120$^{\circ}$ and 170$^{\circ}$. At each of these 5 angles, the scattered light is spectrally and temporally resolved using an imaging spectrometer coupled to a streak camera. 
A multiplexing system is used to record on a single streak camera the scattered beams collected at 25(35)$^\circ$ and at 40(50)$^\circ$, allowing a direct comparison of the emissions at these two angles. A similar configuration is used to combine the signals emitted at 80(90)$^\circ$ and at 110(120)$^\circ$. The interaction region is imaged on the entrance slit of each spectrometer. The probed volume is $\sim 90\times 90\times300~\mu$m$^3$, except for the 180$^{\circ}$ diagnostic for which it was $\sim 300\times300\times500~\mu$m$^3$. For all diagnostics, the covered spectral range is $600-1000$ nm, the spectral and temporal resolutions are $\sim 5$ nm and $\sim 0.1$ ns. These diagnostics are insensitive to the polarization of the collected light.

Stimulated Raman Scattering satisfies the energy conservation, $\omega_0 = \omega_p + \wSRS$, and the momentum conservation, $\vkO = \vkp + \vkSRS$. Here the subscripts 0, p, SRS correspond to the incident laser, the plasma wave and the Raman scattered wave respectively. The plasma wave satisfies the dispersion relation $\omega_p = \omega_{pe}(1+3k_p^2\lambda_D^2)^{1/2}$, where $\omega_{pe} = (n_e e^2 / m_e \epsilon_0)^{1/2}$ is the electron plasma frequency and $\lambda_D$ the Debye length. Since electromagnetic (EM) waves can only propagate if their frequencies are greater than the plasma frequency, $\omega > \omega_{pe}$, SRS can only exist for $\omega_{pe} < \omega_0/2$, or $n_e  < n_c/4$.
Similarly, the stimulated Raman rescattering process should also satisfies the conservation of energy and momentum, $\wSRS = \omega_{p}' + \wReSRS$ and $\vkSRS = \vkpRe + \vkReSRS$, where $\omega_{p}'$ and $\vkpRe$ are the frequency and wavenumber of the plasma wave associated with the rescattered SRS, respectively. The rescattered EM wave has thus a frequency $\wReSRS \sim \omega_0 - \omega_p - \omega_{p}'$, and the rescattering process can only occur in plasma regions where $\wReSRS > \omega_{pe}$.

\begin{figure*}[t]
	\includegraphics[width=\textwidth]{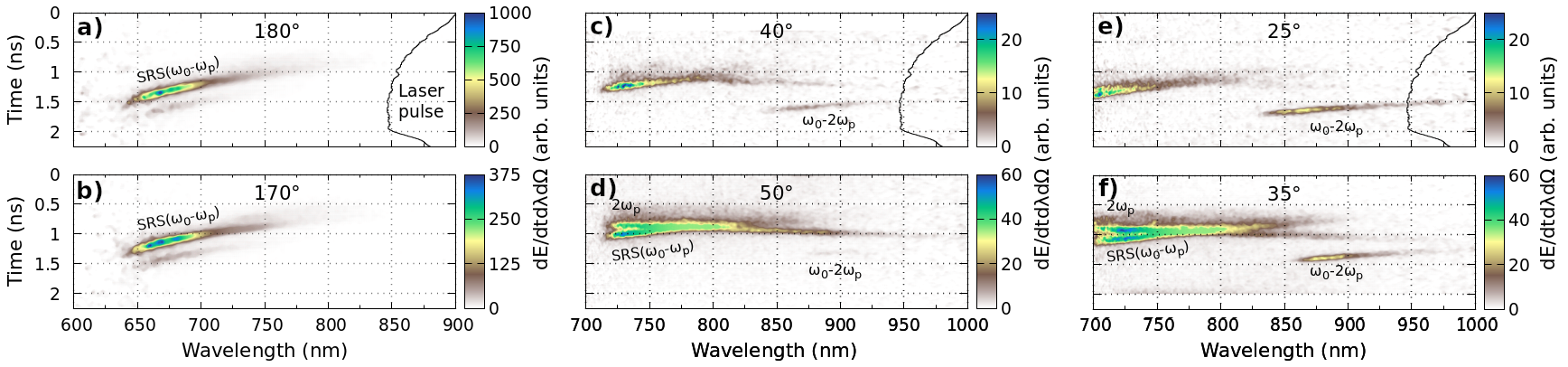}
	\caption{\small{\textit{Time-resolved spectra measured at different angles from the laser axis, for two similar shots, one with the "North"-beam (top panel) and one with the "South"-beam (bottom panel). The black curves show the temporal profile of the laser. The backing pressure is 300 bars, leading to an initial electron plasma density of $\sim 1.2\times 10^{21}$ cm$^{-3}$ ($n_e/n_c \sim$ 0.3). Each spectrum is corrected from the spectral response of the diagnostic. In a) and b) the lower intensity replica 250 ps after the main spectrum is an experimental artifact. In c) and d) the cut at $\sim$ 720 nm comes from a detection limit.}}}
	\label{Spectres_180_45_30_degres}
\end{figure*}

Typical results of the time-resolved frequency spectra of the scattered light measured at different angles are shown in Figure \ref{Spectres_180_45_30_degres} for two similar shots, one with the "North"-beam (top plots) and one with the "South"-beam (bottom plots). The initial plasma density is $n_e/n_c \sim$ 0.3. The left graphs show stimulated Raman backward scattering (SRBS). The H$_2$ gas jet is fully ionized at the very beginning of the laser pulse. The plasma is then heated by inverse bremsstrahlung and starts to expand. When $n_e$ becomes less than $n_c/4$, Raman instability can start to grow. While the plasma is expanding, the plasma frequency decreases, and the Raman scattered light shifts towards the short wavelengths. The Raman signal starts to be detected at plasma densities lower than $n_c/4$, typically below $\sim$ 0.15 $n_c$ ($\lambda <$ 850 nm).
This corresponds to the so-called Raman gap, induced by the collisional damping of the backscattered light~\cite{Barth2024}. At the other angles, three signals are observed: i) the stimulated Raman side scattering (SRSS), with a frequency evolution very close to that of the SRBS; ii) a signal with a frequency that shifts in the opposite direction, from the short to the long wavelengths. It corresponds to the EM emission at $2\omega_p$: the coalescence of two EPWs (Langmuir waves), observed by our group in similar laser-plasma conditions~\cite{Marques2020,Perez2020}; iii) the third signal has a frequency evolution similar to SRBS and SRSS, but shifted towards the infrared side, as expected from stimulated Raman rescattering, with a frequency close to $\omega_0 - 2\omega_p$. The intensity of this signal is greater near the laser axis (25$^\circ$), becomes weaker at larger angles (50$^\circ$), and is absent at 80(90)$^\circ$ or 170(180)$^\circ$. This could be expected from Re-SRBS, the backward rescattering of backward Raman, leading finally to forward scattering. Since the growth rate of SRS is maximum for backward scattering, the rescattering of backward Raman should also be the most intense secondary Raman process, i.e. the one most likely to be detected. Note also that the end of this third signal, at $t \sim 1.5$ ns, coincides with the end of the SRS measured at 170-180$^\circ$ (Fig. \ref{Spectres_180_45_30_degres}-a and -b).

\begin{figure}
	\begin{center}
		\includegraphics[width=\columnwidth]{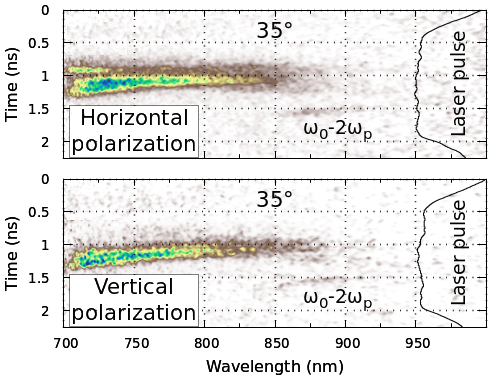}
		\caption{\small{\textit{Time-resolved spectra similar to that in Fig. \ref{Spectres_180_45_30_degres}f, with the addition of a thin polarizer placed in front of the spectrometer, transmitting either horizontal or vertical polarization.}}}
		\label{Spectres_vs_polarization}
	\end{center}
\end{figure}

The emissions were further characterized by their polarization. A thin polarizer was placed in front of the spectrometer to select either horizontal or vertical polarization. Results for both orientations are shown in Fig. \ref{Spectres_vs_polarization}. As expected and already observed~\cite{Marques2020}, the $2\omega_p$ EM emission is polarized in the plane of observation (horizontal). The other two signals are observed for both orientations of the polarizer. This is expected from Raman scattering processes which mainly follows the polarization of the laser, linear at an angle of 45$^{\circ}$ with respect to the horizontal plane.

In the forward direction (Fig. \ref{Spectres_180_45_30_degres}c-f and Fig. \ref{Spectres_vs_polarization}), at early times, the plasma density is too high for the Re-SRBS light to enter the detection window (700-1000 nm), and only the SRS and $2\omega_p$ emissions are observed. As the plasma expands and cools, the plasma frequency decreases, the SRS and $2\omega_p$ signals shift in wavelength and leave the detection window at $t \sim$ 1.2-1.3 ns, the SRS to the short wavelength side ($<$ 720 nm) and the $2\omega_p$ to the long wavelength side ($>$ 1000 nm). At that time, the wavelength of the Re-SRBS signal has decreased enough to enter the detection window, and is observed until $t\sim$ 1.5-1.6 ns. Its disappearance occurs in the middle of the detection range (close to 850 nm), indicating that it is only induced by the end of the rescattering process. This is also correlated with the end of the SRBS (left plots in Fig. \ref{Spectres_180_45_30_degres}-a and -b). Note that both SRBS and Re-SRBS disappear before the end of the laser (see black curve on the right side of the plots in Fig. \ref{Spectres_180_45_30_degres} and Fig. \ref{Spectres_vs_polarization})).

\begin{figure}
	\begin{center}
		\includegraphics[width=\columnwidth]{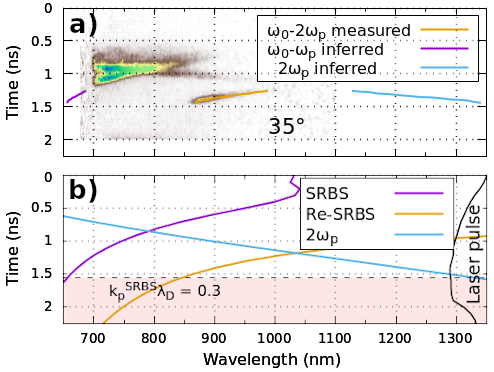}
		\caption{\small{\textit{a): same as Fig. \ref{Spectres_180_45_30_degres}-f but with the SRS and $2\omega_p$ signals at late times ($t>1.2$ ns) inferred from the Re-SRBS signal ($\omega_0-2\omega_p(t)$). b): wavelength evolution of the EM emissions calculated from hydrodynamics simulations.}}}
		\label{Lambda_extrapolated}
	\end{center}
\end{figure}

To cross-check the origin of the signals and their correlation, assuming $\wReSRS = \omega_0 - 2 \omega_p$, we retrieved the plasma frequency from the Re-SRBS signal as it enters the detection window and extrapolated the wavelength of the SRS and $2\omega_p$ signals at late times ($t > 1.2$ ns in Fig. \ref{Spectres_180_45_30_degres}-f). This is illustrated in Fig. \ref{Lambda_extrapolated}-a). Both the SRS and $2\omega_p$ extrapolated signals are in exact continuity with their respective signals measured at earlier times. This confirms that $\wReSRS \sim \omega_0 - 2\omega_p$, meaning that the primary and secondary Raman processes occur at the same plasma density, $\omega_{p}' \sim \omega_p$ (within 5 $\%$, corresponding to the spectral width of the Re-SRBS signal, 35 nm. In addition, the inferred SRS signal (purple curve) ends at $t\sim$ 1.5 ns ($\lambda \sim$ 650 nm), which coincides with the end of the SRS measured at 170-180$^\circ$ (Fig. \ref{Spectres_180_45_30_degres}-a and -b).

The Raman instability can be inhibited by Landau damping if $k_p\lambda_D > 0.3$~\cite{LaFontaine1992}. The values of $k_p$ and $\lambda_D$ evolve with the electron plasma density $n_e$ and temperature $T_e$, which cannot be obtained directly from the spectra in Fig. \ref{Spectres_180_45_30_degres}. To estimate their evolution, two-dimensional (2D) simulations were performed with the radiation hydrodynamics code TROLL\cite{TROLL} (See the supplementary material (SM)), using the measured laser parameters (energy, focal spot and temporal profiles) and plasma parameters (initial density profile) as inputs. The $n_e$, $T_e$ values in the interaction region are then introduced into the dispersion laws and the energy and momentum conservation laws to calculate the evolution of the frequency and wavenumber of the different waves involved (EPW, SRS, Re-SRS and $2\omega_p$ EM waves). Figure \ref{Lambda_extrapolated}-b) shows the calculated wavelength evolution of the EM emissions under the same conditions as in figure \ref{Lambda_extrapolated}-a) and taking $\wReSRS = \omega_0 - 2\omega_p$. It shows very good agreement for the three emissions. In addition, the simulation shows that $k_p\lambda_D$ becomes greater than 0.3 at $t > 1.55$ ns ($n_e/n_c \sim$ 0.03, $T_e \sim$ 465 eV), which means that the Landau damping becomes strong. This agree with the observation: before the end of the laser pulse, the primary SRS is inhibited, which also terminates the rescattering process. Note that for the Re-SRBS plasma wave, $k_p'\lambda_D \sim$ 0.22 at $t\sim$ 1.55 ns, and would become $> 0.3$ at $t>$ 1.9 ns ($n_e/n_c \sim$ 0.019, $T_e \sim$ 405 eV).

\begin{figure}
	\begin{center}
		\includegraphics[width=\columnwidth]{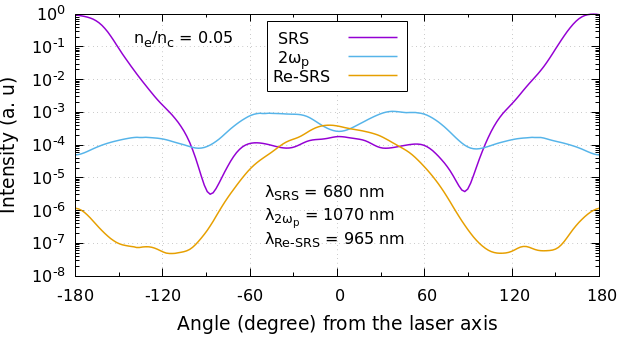}
		\caption{\small{\textit{Angular distribution of the SRS, $2\omega_p$ and $\omega_0-2\omega_p$ signals obtained from Smilei PIC simulations. The amplitudes are normalized to the SRS maximum.}}}
		\label{Smilei_Distrib_angle_signals}
	\end{center}
\end{figure}

To verify that rescattering of the Raman instability is expected in these experimental conditions, we have performed 2D PIC simulations with the Smilei~\cite{Smilei} code. For the initial state of the plasma we use the same hydrodynamic simulations as in Fig. \ref{Lambda_extrapolated}-b). Longitudinal profiles of density and temperature along the central axis are extracted at $t$ = 1.2 ns where, at the focal plane, $n_e/n_c \sim$ 0.051, $T_e\sim$ 500 eV, $T_i\sim$ 200 eV. Since we have observed\cite{Perez2020} that the LPIs are mainly determined by the most intense parts of the focal spot, the size of the simulation domain is defined to include one speckle, with a radius of 4.2 $\mu$m and a maximum intensity of 10$^{15}$ W/cm$^2$ (normalized vector potential of $a_0$ = 0.014). This is the typical intensity of the most intense speckles within the global 500 µm focal spot  shown in Fig. \ref{Setup}, with an average intensity of a few $10^{14}$ W/cm$^2$. The interaction is simulated for up to 20 ps, which is longer than the time required for the Raman and LDI instabilities to grow and saturate\cite{Perez2020}. See the supplementary material (SM) for details.
The angular distributions of the SRS, $2\omega_p$ and $\omega_0-2\omega_p$ EM emissions are shown in figure \ref{Smilei_Distrib_angle_signals}. The SRS and $2\omega_p$ emissions from 20$^\circ$ to 60$^\circ$ have a weak angular dependence, while the $\omega_0-2\omega_p$ signal (Re-SRBS) has a much more pronounced angular dependence. Figure \ref{Smilei_Distrib_angle_signals} also explains why the Re-SRBS was not observed at 80(90)$^\circ$: at this angle the $2\omega_p$ signal is over 200 times larger than the Re-SRBS, preventing its detection. Similarly, at 170(180)$^\circ$ the backward Raman dominates both signals by several orders of magnitude. All these characteristics are in very good agreement with the experimental results shown in Fig. \ref{Spectres_180_45_30_degres}. Note that the relative amplitudes of the forward emitted signals cannot be compared with the experimental data because at $n_e/n_c \sim$ 0.05 ($t\sim$ 1.2 ns in Fig. \ref{Lambda_extrapolated}) the wavelengths of the SRS ($\lambda \sim$ 695 nm) and $2\omega_p$ ($\lambda \sim$ 1090 nm) signals lie outside the detection window.

These PIC simulations also indicate that the intensity of the Raman backscattered light, i.e. the pump for the rescattering Raman, is between few $\%$ to few tens of $\%$ of the laser intensity. Combining these values with the $n_e$ and $T_e$ given by the hydrodynamic simulations, the SRBS and Re-SRBS growth rates have been estimated (see SM), indicating that both Raman processes occur in the absolute regime~\cite{Pesme1973,Michel}, with critical lengths of the order or shorter than the length of the speckles. Under such conditions the primary and secondary processes can occur in very close regions, so that $\omega_p \sim \omega_{p}'$, resulting in the observed rescattered signal at $\wReSRS \sim \omega_0 - 2\omega_p$.

In conclusion, we provide the first experimental evidence of the process of Raman rescattering. The EM emissions produced by SRS, Re-SRBS and Langmuir wave coalescence (EM radiation at $2\omega_p$) have been characterized in terms of frequency, angular distribution and polarization. The time evolution of their frequencies, related to plasma expansion and cooling, is well reproduced by hydrodynamic simulations. The rescattered Raman light ends before the end of the laser, when the primary Raman instability is mitigated by Landau damping ($k_p\lambda_D > 0.3$). The relative amplitudes and angular distributions of the three emissions are in agreement with kinetic (PIC) simulations. This work suggests that Re-SRBS could be a common process in plasmas of moderate density scale lengths and  average laser intensities $\sim 10^{14}$ W/cm$^2$ (with individual speckle intensities close to $10^{15}$ W/cm$^2$), such as those commonly used in direct-drive ICF. The rescattering Raman process can also lead to an underestimation of the primary Raman level in experiments, especially with high intensity short laser pulses. By redistributing the energy, these re-scattering processes (Raman but also Brillouin) may alter the interpretation of experimental diagnostics of the energy balance, including those measuring energetic electrons: while backward and forward SRS can produce energetic electrons in the forward direction, the EPW produced by Re-SRBS will tend to accelerate electrons in the backward direction.

\begin{acknowledgments}
The authors thank the invaluable support of the LULI2000 staff, and the development team of the PIC code Smilei for their help. Smilei simulations were performed with computing and storage resources by GENCI at CINES thanks to the grant 2024-A0170507678 on the supercomputer Adastra's MI250x partition.
\end{acknowledgments}

\bibliography{bibliography}

\end{document}


\title{Experimental evidence of stimulated Raman re-scattering in laser-plasma interaction\\Supplementary material}

\author{J.-R. Marquès}
\email[]{jean-raphael.marques@polytechnique.fr}
\affiliation{LULI, CNRS, CEA, Sorbonne Université, École Polytechnique, Institut Polytechnique de Paris, 91128 Palaiseau, France}

\author{F. Pérez}
\affiliation{LULI, CNRS, CEA, Sorbonne Université, École Polytechnique, Institut Polytechnique de Paris, 91128 Palaiseau, France}

\author{P. Loiseau}
\affiliation{CEA, DAM, DIF, F-91297 Arpajon, France}
\affiliation{Université Paris-Saclay, CEA, LMCE, 91680 Bruyères-le-Châtel, France}

\author{L.~Lancia}
\affiliation{LULI, CNRS, CEA, Sorbonne Université, École Polytechnique, Institut Polytechnique de Paris, 91128 Palaiseau, France}

\author{C. Briand}
\affiliation{LESIA, Observatoire de Paris-PSL, CNRS, Sorbonne Université, Université de Paris, 92195 Meudon, France}

\author{S. Depierreux}
\affiliation{CEA, DAM, DIF, F-91297 Arpajon, France}

\author{M. Grech}
\affiliation{LULI, CNRS, CEA, Sorbonne Université, École Polytechnique, Institut Polytechnique de Paris, 91128 Palaiseau, France}

\author{C. Riconda}
\affiliation{Sorbonne Université, LULI, CNRS, CEA, École Polytechnique, Institut Polytechnique de Paris, 75255 Paris, France}

\maketitle

\subsection{Hydrodynamic simulations}

Plasma conditions are calculated using the radiation hydrodynamics code TROLL\cite{TROLL}. TROLL is a multi-dimensional arbitrary Lagrangian-Eulerian code allowing planar and cylindrical geometries, and uses a ray-tracing package modeling the laser propagation and absorption. The experiment is simulated in two-dimensional Lagrangian planar geometry, on 1 $mm^2$ uniform grid. We use experimentally measured profiles of the gas jet to initialize the plasma density, with uniform ion and electron temperatures. We use also the experimental laser pulse power. The hydrogen gas jet is then heated by inverse Bremsstrahlung absorption mechanism, and electron and ion temperatures spread due to Spitzer-Härm thermal conductivity.

\begin{figure}[h!]
	\begin{center}
		\includegraphics[width=0.8\columnwidth]{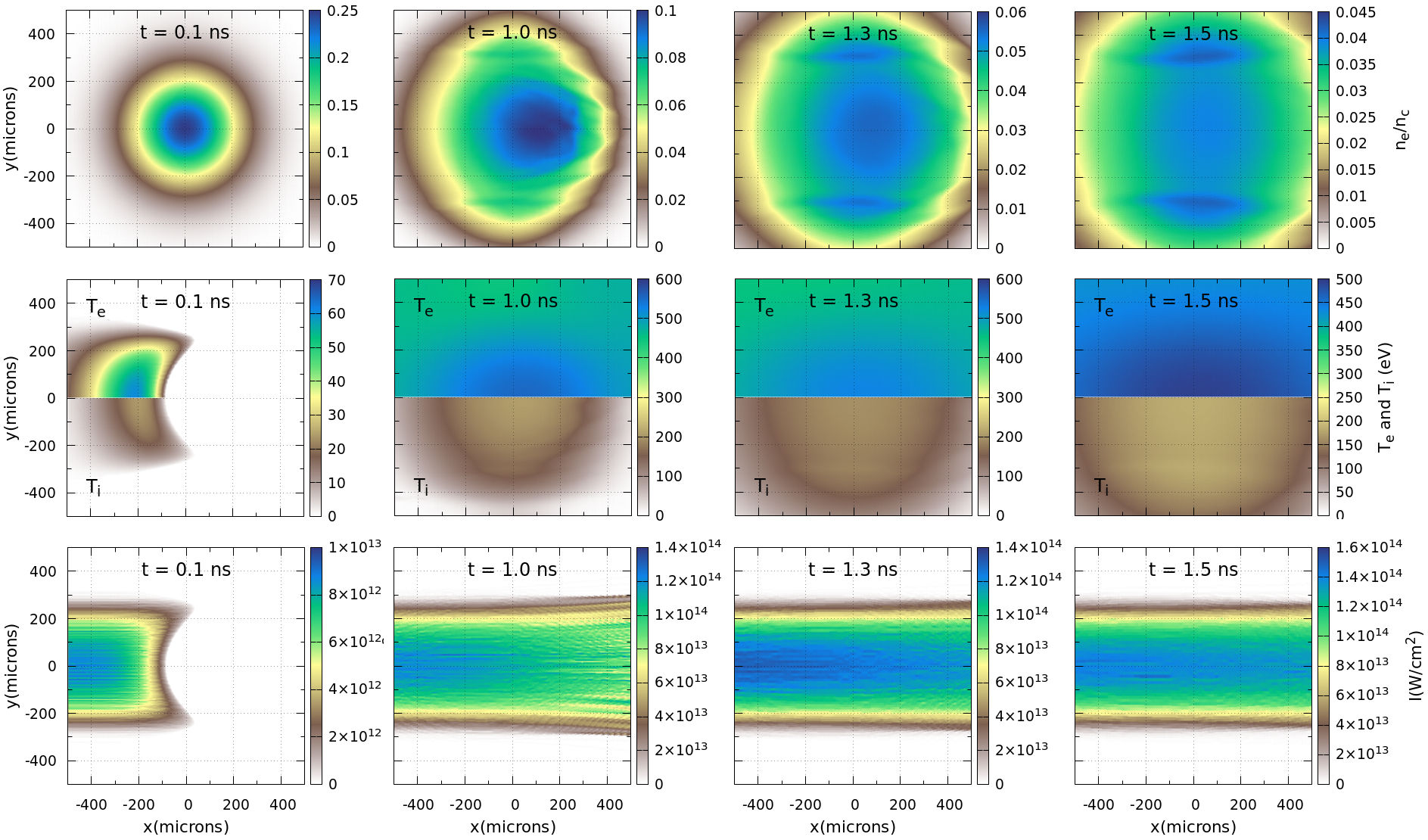}
		\caption{\small{\textit{Spatial profiles of the plasma density ($n_e/n_c$, top panels), temperatures ($T_e$ and $T_i$, middle panels), and laser intensity (bottom panels) at four different times in the laser pulse (See Fig. \ref{Ne_Te_Ti_Ilaser_vs_time} for the pulse temporal profile).}}}
		\label{Profil2D_Ne_Te_Ti}
	\end{center}
\end{figure}

\begin{figure}[h!]
	\begin{center}
		\includegraphics[width=0.45\columnwidth]{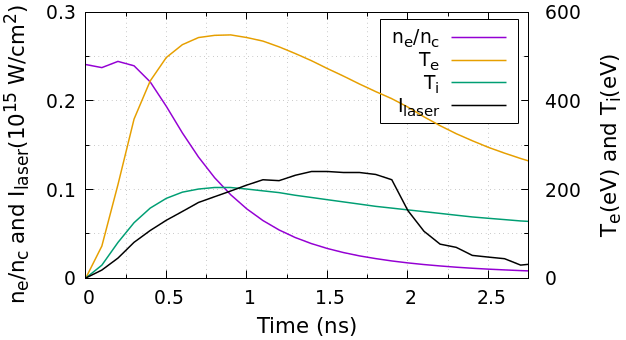}
		\caption{\small{\textit{Temporal evolution of the plasma density ($n_e/n_c$), temperatures ($T_e$ and $T_i$), and laser intensity, from the same simulation as in than Fig. \ref{Profil2D_Ne_Te_Ti}, and extracted from the region probed in the experiment: 50 $\times$ 50 µm at the center ($x$ = 0, $y$ = 0) of the plasma.}}}
		\label{Ne_Te_Ti_Ilaser_vs_time}
	\end{center}
\end{figure}

Figure \ref{Profil2D_Ne_Te_Ti} shows two-dimensional profiles of the density ($n_e/n_c$), temperatures ($T_e$ and $T_i$) and laser intensity at four different times in the laser pulse.  The time evolution of these quantities at the plasma (gas jet) center are presented in figure \ref{Ne_Te_Ti_Ilaser_vs_time}. These extracted $n_e$ and $T_e$ values are then introduced into the dispersion laws and the energy and momentum conservation laws to calculate the evolution of the frequency and wavenumber of the different waves involved (EPW, SRS, Re-SRS and $2\omega_p$ EM waves), which is presented in Fig. 4-b) of the main manuscript.

\begin{figure}[h!]
	\begin{center}
		\includegraphics[width=0.45\columnwidth]{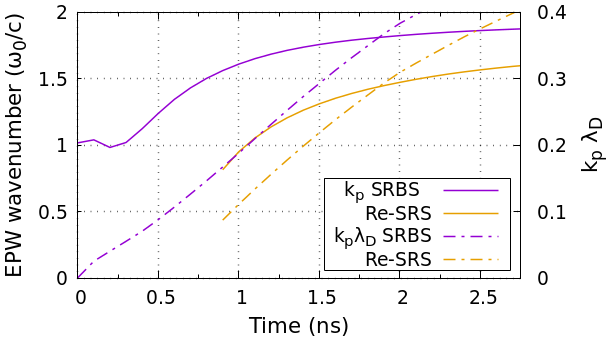}
		\caption{\small{\textit{Temporal evolution of the wavenumber and Landau damping factor of the EPWs excited by the backward SRS and the Re-SRBS that scatters at 30° from the laser axis. Calculated from the $n_e(t)$ and $T_e(t)$ values shown in Fig. \ref{Ne_Te_Ti_Ilaser_vs_time}.}}}
		\label{kp_kplD_vs_time}
	\end{center}
\end{figure}

Figure \ref{kp_kplD_vs_time} shows the temporal evolution of the wavenumbers and the Landau damping factors of the EPWs excited by the backward SRS and the Re-SRBS. Calculated from the $n_e$ and $T_e$ values presented in Fig. \ref{Ne_Te_Ti_Ilaser_vs_time} and for the rescattering process which scatters light at 30$^\circ$ from the laser axis, as observed experimentally (Fig. 2e and f in the main manuscript)

\begin{figure}[h!]
	\begin{center}
		\includegraphics[width=0.45\columnwidth]{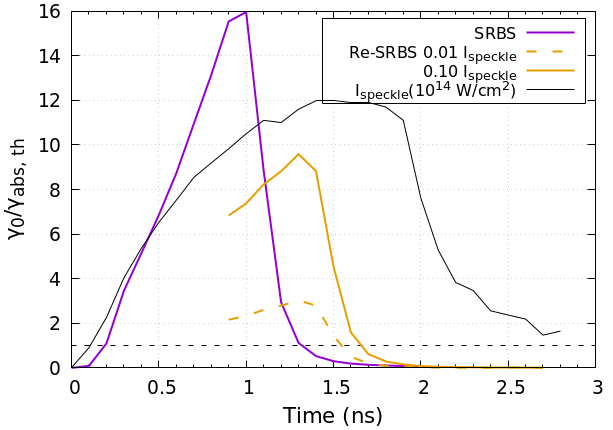}
		\caption{\small{\textit{SRBS and Re-SRBS growth rates $\gamma_0$ normalized to their respective absolute growth rates $\gamma_{abs, th}$. The laser intensity in the speckle is 10 times the spatially averaged focal spot intensity used in the hydrodynamic simulation of Fig. \ref{Ne_Te_Ti_Ilaser_vs_time}.}}}
		\label{SBS_absolute_threshold}
	\end{center}
\end{figure}

Figure \ref{SBS_absolute_threshold} shows the SRBS and Re-SRBS growth rates normalized to their respective absolute growth rates \cite{Michel}-(, chapter 6.6.3), calculated from the $n_e$ and $T_e$ values presented in Fig. \ref{Ne_Te_Ti_Ilaser_vs_time} and, in the case of Re-SRBS, for two values of the back-scattered SRBS light, 1 $\%$ and 10 $\%$ of the incident laser intensity in the speckles, in accordance with the PIC simulations (see below). Both the SRBS and Re-SRBS growth rate are significantly above the threshold for a growth in the absolute mode.

\subsection{Particle-In-Cell simulations}

The particle-in-cell simulations were carried out using the code Smilei v5.1 \cite{Smilei}. The two-dimensional simulation domain was a 42$\times$17 $\mu$m$^2$ rectangle with a cell size of $\sim 13\times13$ nm, and the time step was set to $27\times 10^{-6}$ ps. To match hydrodynamics simulations, the plasma was made of two species, hydrogen ions and electrons, both with a charge density gradient ranging from $0.050\,n_c$ at the laser entrance to $0.052\,n_c$ at the exit of the simulation box. The electron and ion temperatures were set to 500 and 200 eV, respectively. These two species corresponded to a total of 2.2 billion macro-particles. The laser of wavelength 527 nm had a constant intensity, and a Gaussian transverse profile with a full width at half maximum of 5 $\mu$m.

These simulations were run on 32 graphical processing units (model MI250 from AMD) for 48 hours, in the Adastra supercomputer from CINES [https://www.cines.fr/calcul/adastra].

\begin{figure}
	\begin{center}
		\includegraphics[width=0.4\columnwidth]{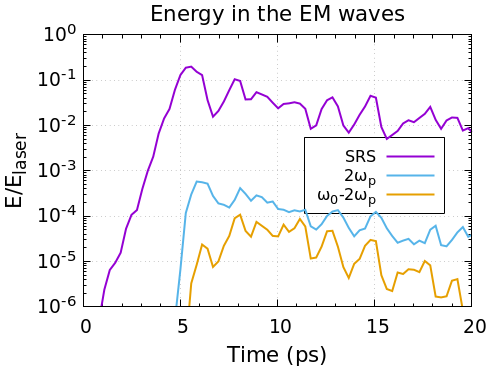}
		\caption{\small{\textit{Temporal evolution of the energy of the electromagnetic waves, normalized to the laser energy.}}}
		\label{Energies_vs_time}
	\end{center}
\end{figure}

Figure \ref{Energies_vs_time} shows the temporal evolution of the energy of the electromagnetic waves, normalized to the laser energy. The primary Raman instability grows and is closely followed by the growth of the Raman rescattering as well as the Langmuir wave coalescence processes, producing EM waves at $\omega_0-2\omega_p$ and $2\omega_p$ respectively. These processes saturate at $\sim$ 5 ps, before the end of the simulation, at 20 ps.

\begin{figure}[t]
	\begin{center}
		\includegraphics[width=0.8\columnwidth]{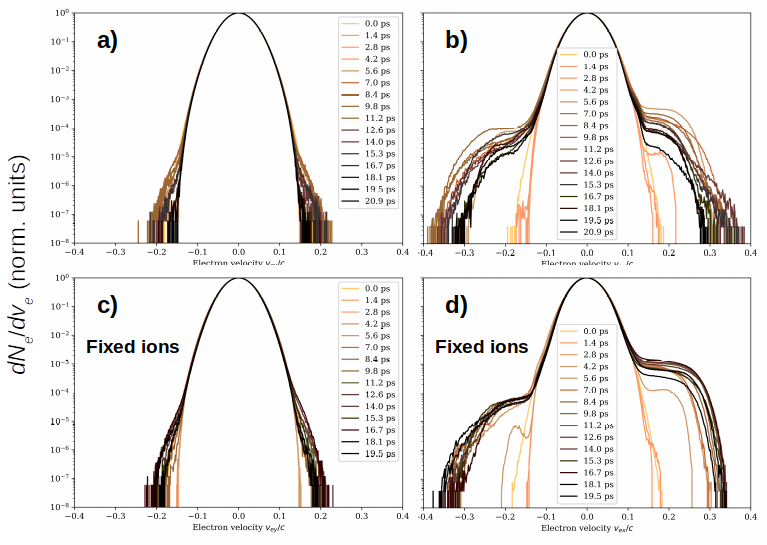}
		\caption{\small{\textit{Electron velocity distribution in the direction perpendicular ($y$, a) and c)) and along the laser axis ($x$, b) and d)). Simulations in c) and d) are for the same laser-plasma conditions than a) and b), but with fixed ions.}}}
		\label{Distrib_vitesses}
	\end{center}
\end{figure}

Figure \ref{Distrib_vitesses} shows the velocity distributions of the electrons along the laser axis or perpendicularly, from two types of simulations: with moving ions (a and b) or fixed ions (c and d). In both cases, the acceleration is mainly along the laser axis, with similar velocities in the forward and backward directions, up to $\sim 0.35c$ ($\sim$ 35 keV). The electron plasma waves associated with the forward (SRFS) or backward Raman (SRBS) have both a phase velocity in the direction of the laser and are the main source of forward acceleration. The dominant Raman rescattering process is that associated with backward Raman. Re-SRBS produces a plasma wave with a backward phase velocity and therefore potentially at the origin of the backward hot electrons. However, the Langmuir decay instability (LDI) also produces EPWs with backward phase velocity. To determine the contribution of these two processes, simulations were performed with fixed ions, thus preventing the excitation of the LDI. The results are shown in figure \ref{Distrib_vitesses}c and d. The increased number of forward and backward hot electrons produced at late times is due to the fact that without LDI the EPWs excited by SRBS or SRFS last longer. Despite this overall increase, the energy and the number of backward hot electrons remain almost unchanged, confirming that it is the Re-SRBS that produces the backward hot electrons.

\bibliography{bibliography_SM}